\documentclass{article}

\usepackage{perseus}
\usepackage[utf8]{inputenc}
\usepackage[T1]{fontenc}
\usepackage{microtype}
\usepackage{graphicx}
\usepackage{booktabs}
\usepackage{caption}
\usepackage{nicefrac}
\usepackage{pifont}
\usepackage{ragged2e}
\usepackage{xspace}
\usepackage{threeparttable}
\usepackage{nicematrix}
\makeatletter
\@ifpackagelater{nicematrix}{2026/02/22}{}{%
  \AddToHook{env/threeparttable/begin}{%
    \TPT@hookin{NiceTabular}\TPT@hookin{NiceTabular*}\TPT@hookin{NiceTabularX}}}
\makeatother
\usepackage{tabularx}
\usepackage{fontawesome5}
\usepackage[style=numeric-comp, sortcites=true, sorting=none, labeldateparts=true, backend=biber, maxbibnames=10]{biblatex}
\usepackage[hyperfootnotes=false]{hyperref}
\usepackage[capitalize,noabbrev]{cleveref}
\newcommand{\circled}[2][0.8]{%
    \tikz[baseline=(char.base), scale=#1]{
        \node[
            circle,
            draw,
            line width=0.6pt,
            inner sep=1.5pt,
            font=\footnotesize\bfseries,
            transform shape,
        ] (char) {#2};
    }%
}

\definecolor{nicecitblue}{rgb}{0.21,0.49,0.74}
\hypersetup{colorlinks, allcolors=nicecitblue}
\venue{Preprint.}

\newcommand{\repo}{https://github.com/gladia-research-group/perseus}
\newcommand{\speedrunurl}{https://github.com/gladia-research-group/perseus/tree/speedrun}

\newcommand{\stok}{\,\nicefrac{\textrm{s}}{\textrm{tok}}}

\newcommand{\yes}{\textcolor{green!55!black}{\ding{51}}}
\newcommand{\no}{\textcolor{red!70!black}{\ding{55}}}

\DeclareCiteCommand{\tabcite}
  {\usebibmacro{prenote}}
  {\bibhyperref{\printnames{labelname}, \printlabeldateextra}}
  {\multicitedelim}
  {\usebibmacro{postnote}}

\NiceMatrixOptions{custom-line = {command = dashedline, tikz = dashed, total-width = \pgflinewidth}}

\newcommand{\llama}{ \circled{1}\xspace}
\newcommand{\gpt}{ \circled{2}\xspace}
\newcommand{\vit}{ \circled{3}\xspace}
\newcommand{\bert}{ \circled{4}\xspace}

\begin{document}

\begin{titlebox}
  {\fontsize{17pt}{21pt}\selectfont\bfseries\color{black!85}
    Perseus: A Full-Stack System for Homomorphic\\ Model Inference\par}
  \vspace{10pt}
  {\normalsize\color{black!85}
    Alessandro Zirilli$^{\star 1}$\qquad
    Davide Marincione$^{\star 1,3}$\\[3pt]
    Evgenios M. Kornaropoulos$^{2}$\qquad
    Giuseppe Ateniese$^{2}$\qquad
    Emanuele Rodol\`a$^{1,3}$\par}
  \vspace{6pt}
  {\small\color{black!60}
    $^1$Sapienza University of Rome\qquad
    $^2$George Mason University\qquad
    $^3$Paradigma\par}
  \vspace{3pt}
  \vspace{14pt}

  \begin{minipage}{0.86\linewidth}
    \small\justifying
    Large Language Models ({\scshape LLM}s) have made it exceptionally convenient for users to send their private data to model providers, which process them in plaintext. Fully Homomorphic Encryption ({\scshape FHE}) promises to secure remote inference by computing directly on encrypted data, yet it has remained prohibitively slow. The effort to make it practical remains fragmented, with most existing frameworks being \emph{closed-source}, which keeps progress from compounding.
    We present \texttt{Perseus}, a free and \emph{open}$^1$ Python software library that supports the full stack of encrypted model inference, from the cryptographic backend and data packing, to encrypted nonlinearities and end-to-end model execution.
    \texttt{Perseus} ships with built-in novel optimizations, one at each level of the stack: in-kernel regeneration of key-switching keys at the (i) backend level, a complex variant of \texttt{Cachemir} packing~\cite{cachemir} at the (ii) data level, and a state-of-the-art bootstrap placer that is both packing- and magnitude-aware at the (iii) model level.
    \texttt{Perseus} achieves state-of-the-art latency across multiple models and modalities, within $3\%$ of the plaintext models' accuracy and perplexity. Remarkably, it is $3.96\times$ faster at single-token decoding than any other implementation on \texttt{Llama-3-8B} (\cref{tab:decode}). We initiate, for the first time, a public speedrun on encrypted \texttt{GPT-2}, so that progress can compound.
  \end{minipage}

  \begin{minipage}{\linewidth}
    \centering
    \small
    \setlength{\tabcolsep}{5pt}
    \begin{threeparttable}
    \captionof{table}{\justifying Encrypted \texttt{Llama-3-8B} decode of one token; $^\dagger$composed from per-block or per-operator timings, not run end to end. Throughout the paper we mark competing frameworks as: \yes\, available, $\sim$ temporarily or partially released, and \no\, closed-source.}
    \begin{NiceTabular}{lrrlc}
      \toprule
      \llama\textbf{\texttt{Llama-3-8B}} & \textbf{context} & \textbf{s$/$tok} & \textbf{{\scshape GPU}} & \textbf{availability} \\
      \midrule
      \texttt{AR-HE} (\tabcite{arhe})        & $32$  & $1569.0\rlap{$^{\dagger}$}$ & {\scshape H100}         & \no \\
      \texttt{ROSETTA} (\tabcite{rosetta})   & $512$ & $311.0\rlap{$^{\dagger}$}$  & {\scshape A100}         & \no \\
      \texttt{Cachemir} (\tabcite{cachemir}) & $512$ & $97.0\rlap{$^{\dagger}$}$   & {\scshape A100}         & $\sim$ \\
      \texttt{Sylph} (\tabcite{sylph})       & $128$ & $31.0\rlap{$^{\dagger}$}$                      & {\scshape RTX PRO 6000} & \no \\
      \rowcolor{RoyalBlue!12}\textbf{\texttt{Perseus} (ours)} & $\mathbf{512}$ & $\mathbf{7.8}$ & \textbf{{\scshape RTX PRO 6000}} & \yes \\
      \bottomrule
    \end{NiceTabular}
    \label{tab:decode}
    \end{threeparttable}
  \end{minipage}

  \vspace{12pt}
  \raisebox{-0.2\height}{\includegraphics[width=1em,height=1em]{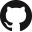}}\hspace{0.3em}%
  \href{\repo}{\small\texttt{Code}}\hspace{1.5em}%
  \faFlagCheckered\hspace{0.3em}%
  \href{\speedrunurl}{\small\texttt{Speedrun}}
\end{titlebox}

{\renewcommand{\thefootnote}{}%
 \footnotetext{$^{\star}$Equal contribution. Contact: \texttt{\{zirilli, marincione\}@di.uniroma1.it}.}
 \noindent
  \footnotetext{$^1$Released under the BUSL 1.1: free for research and non-commercial use, commercial use available upon request.}}

\captionsetup{font=small, skip=6pt, position=below, hypcap=false,
              justification=justified, singlelinecheck=false}
\section{Introduction}
Fully Homomorphic Encryption ({\scshape FHE}) promises a future of trustless outsourced computation. One of its most appealing applications is Machine Learning, where successful task execution often depends on disclosing sensitive data. In particular, the recent expansion of capabilities in Large Language Models ({\scshape LLM}s) has made them extremely convenient for many users, who accept to share their private, plaintext data with the models' inference providers. While encrypted models exist, the extreme latency {\scshape FHE} introduces through bootstrapping 
limits the deployment of these models in practice.

The landscape of encrypted execution is evolving rapidly, with various approaches aiming to reduce the high latency imposed by {\scshape FHE} (and in particular {\scshape CKKS}~\cite{ckks}). 
New techniques are being released for all the different abstraction levels of the encrypted models' inference stack, from bootstrap speed-ups, {\scshape GPU} implementations~\cite{fideslib,cheddar,theodosian,heaan}, and $32$-bit arithmetic~\cite{cheddar} at the cryptographic backend level; to task-specific packing systems~\cite{cachemir,moai,nexus,stip,tricycle,arion,themis,euston} at the data level; and to new techniques for approximating machine learning layers~\cite{thor,gaussiankernel,powersoftmax,sylph,cutmax,encinvsqrt,encsign,powerformer,fiona}, bootstrap placers~\cite{dacapo,fhelipe,orion}, and fine-tuning techniques~\cite{heat,powerformer} at the model level. The unfortunate reality of these optimizations is that most are implemented into different, sometimes \emph{closed-source}, frameworks, making integration between one another difficult and time-consuming.

We present \texttt{Perseus}, a \emph{publicly available Python library} for encrypted machine learning model inference that enables practitioners to implement models and optimization techniques in an easy-to-use environment. \texttt{Perseus}' contributions are:

\begin{itemize}
  \item a free and open Python library (\href{\repo}{code}) that supports full-stack development of models, built on \texttt{Perseus-core}, a Rust and {\scshape CUDA} cryptographic backend for {\scshape CKKS} \texttt{32-bit} arithmetic and bootstrapping.
  \item Built-in optimizations at all levels of the stack, from in-kernel regeneration of key-switching keys using \texttt{ChaCha12} (halving key storage), a complex variant of \texttt{Cachemir} packing (roughly halving memory), to packing- and magnitude-aware bootstrap planning with an iterated minimum cut (${\approx}30\%$ fewer bootstraps), which result in state-of-the-art latency on all already-implemented models (\texttt{GPT-2}, \texttt{Llama-3-8B}, \texttt{ViT}, \texttt{BERT}).
  \item A public leaderboard to cooperatively optimize encrypted language models' inference, a speedrun (\href{\speedrunurl}{\small{leaderboard}}) that will condense the community's effort to reduce latency.
\end{itemize}

Throughout the paper, we report published results for competing methods when their code is unavailable and reproduce their results when the code is publicly available.

\section{Perseus}
We conceptualize an encrypted models' inference stack into three abstraction levels: (i) the \emph{cryptographic backend} that executes {\scshape CKKS} primitives (\emph{e.g.}\ encryption, additions, multiplications, bootstrapping, etc.); (ii) the \emph{data level}, which controls the layout of ciphertexts' slots (the packing) and thus defines encrypted algorithms; and (iii) the \emph{bootstrap placement} that decides where and how to refresh the multiplicative depth of the model's circuit. 

State-of-the-art {\scshape CKKS} execution of {\scshape LLM}s is still far from practical, despite frequent optimizations across all levels of the stack. Although many optimizations are orthogonal and could be combined to compound their benefits, doing so remains difficult in practice, as the closed-source nature of most such systems fragments the stack and limits opportunities for integration.

\texttt{Perseus} offers an alternative that supports the entire encrypted inference stack. Adopters can implement a new encrypted machine learning model, a packing layout, a new approximation, and/or a bootstrap placement algorithm in pure Python on top of \texttt{Perseus-core}, \emph{i.e.}, our Rust and CUDA {\scshape CKKS} runtime. In \texttt{Perseus}, the ease of development comes with state-of-the-art end-to-end performance, driven not only by an efficient implementation designed with machine learning models in mind, but also by three contributions, one for each level of the stack.

\paragraph{(i) Cryptographic backend: in-kernel key regeneration.} Key-switching keys dominate {\scshape GPU} memory and bandwidth during bootstrapping. Half of each key is uniformly random, and thus can be regenerated from a seed. Instead of storing this half, \texttt{Perseus} recomputes it with a counter-mode \texttt{ChaCha12} expander inside the two kernels that make use of the key. The remaining, non-random, part is stored as a dense \texttt{28-bit} stream, trading extra arithmetic for less memory traffic and halving key storage.

\paragraph{(ii) Data: complex-\texttt{Cachemir}.} \texttt{Cachemir}~\cite{cachemir} packs activations and key/value caches for autoregressive decoding, but exploits only the real slots of a ciphertext. \texttt{Perseus} extends it to the complex slots of {\scshape CKKS}, carrying two real vectors in each ciphertext, which roughly doubles the throughput of many operations and layers, and roughly halves memory requirements.

\paragraph{(iii) Model: bootstrap placer.} Since bootstraps are two to three orders of magnitude slower than any other primitive, minimizing their usage makes a concrete difference in reducing latency. \texttt{Perseus} first captures a model's computation graph, typed by each ciphertext's coefficient magnitude and packing periodicity, then optimizes the count by placing bootstraps via an iterated minimum cut over this typed graph, unlike prior min-cut placers~\cite{resbm} that decide from depth and scale alone. The annotations about period and magnitude are used by selecting the most appropriate bootstrap for both, imposing a cheaper and more accurate sparse bootstrap when possible, and choosing the right scaling factor to keep coefficients in the \texttt{EvalMod} approximation band. On average, the placer executes ${\approx}30\%$ fewer bootstraps than comparable placers~\cite{orion,fhelipe,dacapo}.

\begin{table}[h]
  \centering
  \small
  \setlength{\tabcolsep}{5pt}
  \begin{threeparttable}
    \caption{Encrypted \texttt{Llama-3-8B} prefill; $^\dagger$composed from per-block or per-operator timings, $^*$amortized over a batch, $^\S$replicated from their released code. $\!\to\!$ linearly scaled timings to adjust prompt size. \yes\, available, \no\, closed-source, $\sim$ temporarily or partially released.}
    \begin{NiceTabularX}{\textwidth}{
      >{\hsize=1.2\hsize\raggedright\arraybackslash}X 
      r
      r
      >{\hsize=0.8\hsize\raggedright\arraybackslash}X
      c
    }
      \toprule
      \llama\texttt{Llama-3-8B} & prompt & prefill (s) & \textbf{{\scshape GPU}} & availability \\
      \midrule
      \texttt{MOAI} (\tabcite{moai})       & $8\!\to\!128$ & $3592.0\rlap{$^{\dagger*}$}$ & {\scshape NVIDIA H200 GPU}  & $\sim$ \\
      \texttt{NEXUS} (\tabcite{nexus})     & $8\!\to\!128$ & $829.0\rlap{$^{\dagger*}$}$  & $4\times${{\scshape A100}} & $\sim$ \\
      \texttt{Euston} (\tabcite{euston})   & $128$         & $484.0\rlap{$^{\dagger*}$}$  & {\scshape RTX 6000 Ada}  & $\sim$ \\
      \texttt{STIP} (\tabcite{stip})       & $128$         & $406.6\rlap{$^{\dagger*}$}$  & {\scshape A100}          & \no \\
      \texttt{Odin} (\tabcite{odin})       & $128$         & $366.4$                      & {\scshape H100}          & $\sim$ \\
      \texttt{Cerium} (\tabcite{cerium})   & $128$         & $347.5\rlap{$^{\S}$}$        & {\scshape RTX PRO 6000}  & \yes \\
      \texttt{Sylph} (\tabcite{sylph})     & $128$         & $116.0\rlap{$^{\dagger}$}$   & {\scshape RTX PRO 6000}  & \no \\
      \rowcolor{RoyalBlue!12}\textbf{\texttt{Perseus} (ours)} & $128$ & $\mathbf{54.8}$ & {\scshape RTX PRO 6000} & \yes \\
      \bottomrule
    \end{NiceTabularX}
    \label{tab:llama-prefill}
  \end{threeparttable}
\end{table}
\section{Results}
All \texttt{Perseus}'s measurements use one {\scshape NVIDIA RTX PRO 6000} and a 32-bit {\scshape RNS} {\scshape CKKS} chain with composite scaling of degree 2 (each level a pair of $27$-bit primes, $q_0$ a pair of $28$-bit primes), $\log N = 16$, 128-bit security and $q_0 = 2^{56}$, giving $27$ levels over $51$ primes.

We implement in \texttt{Perseus} four deep-learning models,\llama\texttt{Llama-3-8B},\gpt\texttt{GPT-2-small},\vit\texttt{ViT-B/16}, and\bert\texttt{BERT-base}, and report their latency end-to-end, \emph{i.e.}\ accounting for weights movement and management during the actual inference. We address two execution modalities, one being \emph{prefilling} (or analogously \emph{classification}) and the other being single-token \emph{decoding}. They fundamentally differ as the first involves multiple tokens being processed at the same time, while the latter autoregressively generates tokens considering the current {\scshape KV}-caches' state.

For prefill and classification we also report the normalized accuracy, \emph{i.e.,} the accuracy of the encrypted model over the plaintext one on the same inputs (next-token accuracy for prefill, label accuracy for classification), as a proxy for performance degradation due to inherent {\scshape CKKS} approximated arithmetic, bootstraps, and required nonlinearities' approximations. For decoding, we report the perplexity of the encrypted model next to the reference plaintext one. All quality metrics are computed on held-out data from the approximations' calibration set.

\llama\texttt{Llama-3-8B} is the most challenging model we test, due to the sheer number of parameters and operations it requires. It is one of the most popular models transferred to FHE in the literature, and, when it becomes practical to run, it will mark a milestone, as it is in the same size class as state-of-the-art models. We test it in both execution modalities, decode (\Cref{tab:decode}) and prefill (\Cref{tab:llama-prefill}). 

 Decode uses in-kernel key regeneration, our complex-\texttt{Cachemir} packing and complex-{\scshape KV}-cache, and \texttt{Cutmax}~\cite{cutmax} argmax. Under a fixed bootstrap schedule, decided with our planner, \texttt{Perseus} reduces latency by $3.96\times$ with respect to the fastest known implementation (\emph{i.e.}\ \texttt{Sylph}~\cite{sylph}, running on roughly the same hardware, but on the closed-source \texttt{HEaaN}~\cite{heaan}). The encrypted perplexity settles on $7.77$ against a plaintext value of $7.76$ (at $512$ context).
\begin{table}[h]
  \centering
  \small
  \setlength{\tabcolsep}{5pt}
  \begin{threeparttable}
    \caption{Encrypted \texttt{GPT-2}: prefill of a prompt and decode of one token; $^\dagger$composed from per-block timings. \yes\, available, \no\, closed-source, $\sim$ temporarily or partially released.}
\begin{NiceTabularX}{\textwidth}{
      >{\raggedright\arraybackslash}p{5em}
      >{\hsize=1.5\hsize\raggedright\arraybackslash}X
      @{\hspace{3pt}}
      r r
      >{\hsize=0.5\hsize\raggedright\arraybackslash}X
      c
    }
      \toprule
      & \gpt\texttt{GPT-2} small & prompt/context & latency & \textbf{{\scshape GPU}} & availability \\
      \midrule
      \Block[l]{2-1}{prefill (s)}
        & \texttt{AR-HE} (\tabcite{arhe})                & $128$ & $18520.0$                 & {\scshape H100} & \no \\
        & \rowcolor{RoyalBlue!12}\textbf{\texttt{Perseus} (ours)} & $128$ & $\mathbf{6.8}$ & \rlap{\scshape RTX PRO 6000} & \yes \\[4pt]
      \dashedline
      \noalign{\vskip 4pt}
      \Block[l]{4-1}{decode (\stok)}
        & \texttt{AR-HE} (\tabcite{arhe})                & $33$  & $544.0$                   & {\scshape H100} & \no \\
        & \texttt{ROSETTA} (\tabcite{rosetta})           & $512$ & $78.6\rlap{$^{\dagger}$}$ & {\scshape A100} & \no \\
        & \texttt{EncryptedLLM} (\tabcite{encryptedllm}) & $128$ & $68.0$                    & {\scshape A100} & $\sim$ \\
        & \rowcolor{RoyalBlue!12}\textbf{\texttt{Perseus} (ours)} & $512$ & $\mathbf{1.9}$ & \rlap{\scshape RTX PRO 6000} & \yes \\
      \bottomrule
    \end{NiceTabularX}
    \label{tab:gpt2}
  \end{threeparttable}
\end{table}
Prefill uses {\scshape PCMM} \cite{thor, secure, pcmm, powerformer, sylph} for the {\scshape MLP} execution algorithm, and {\scshape CCMM} \cite{pcmm, fastbatch, ccmm, sylph} for the Attention mechanism, which are two {\scshape FHE} implementations of regular Generalized Matrix Multiplication (unlike decode, which uses a Baby-Step-Giant-Step algorithm). Analogously to decode, \texttt{Perseus} establishes itself as the fastest implementation. In prefill, normalized accuracy is $100\%$ over $64$ held-out prompts.

\gpt\texttt{GPT-2-small} is a toy model, but at $1.9\stok$ ($41\times$ faster than the closest baseline, \Cref{tab:gpt2}) we believe it can provide a fast-iteration development platform for further {\scshape ML}-dedicated optimizations in the encrypted domain. We therefore publish and commit to maintaining a public \href{\speedrunurl}{leaderboard} that invites the community to improve encrypted \texttt{GPT-2}. Our objective is to trigger a snowball effect similar to what the plaintext community is experiencing~\cite{modded_nanogpt_2024,noise_2026}. Our implementation maintains a perplexity of $26.46$, compared with $26.47$ in plaintext. Its prefill reaches a normalized accuracy of $100\%$ over $64$ held-out prompts.
\begin{table}[h]
  \centering
  \small
  \setlength{\tabcolsep}{5pt}
  \begin{threeparttable}
    \caption{Encrypted \texttt{ViT} inference of one image; $^*$amortized over a batch of $32$, on a model retrained with polynomial layers. \yes\, available, \no\, closed-source, $\sim$ temporarily or partially released.}
    \begin{NiceTabularX}{\textwidth}{
      >{\hsize=1.2\hsize\raggedright\arraybackslash}X 
      l
      r
      >{\hsize=0.8\hsize\raggedright\arraybackslash}X
      c
    }
      \toprule
      \vit\texttt{ViT} & model & s$/$img & \textbf{Hardware} & availability \\
      \midrule
      \texttt{ATLAS} (\tabcite{atlas}) & \texttt{ViT-B/16} & $2259.0$              & {\scshape RTX 4090 48GB} & \yes \\
      \texttt{FIONA} (\tabcite{fiona}) & \texttt{ViT-S/16} & $374.4\rlap{$^{*}$}$ & {\scshape CPU}, 48 cores        & \no \\
      \rowcolor{RoyalBlue!12}\textbf{\texttt{Perseus} (ours)} & \texttt{ViT-B/16} & $\mathbf{21.1}$ & {\scshape RTX PRO 6000} & \yes \\
      \bottomrule
    \end{NiceTabularX}
    \label{tab:vit}
  \end{threeparttable}
\end{table}
\begin{table}[h!]
  \centering
  \small
  \setlength{\tabcolsep}{5pt}
  \begin{threeparttable}
    \caption{Encrypted \texttt{BERT} inference of one input. $^\S$replicated from their released code on our hardware.  \yes\, available, \no\, closed-source, $\sim$ temporarily or partially released.}
    \begin{NiceTabularX}{\textwidth}{
      >{\hsize=1.4\hsize\raggedright\arraybackslash}X 
      l
      r
      r
      >{\hsize=0.6\hsize\raggedright\arraybackslash}X
      c
    }
      \toprule
      \bert\texttt{BERT} & model & tokens & s/sentence & \textbf{{\scshape GPU}} & availability \\
      \midrule
      \texttt{THOR} (\tabcite{thor})    & \texttt{BERT-base} & $128$ & $602.0$ & {\scshape A100} & $\sim$ \\
      \texttt{POLARIS} (\tabcite{sok})  & \texttt{BERT-Mini} & --    & $83.0$  & {\scshape A100} & \no \\
       \texttt{Cerium} (\tabcite{cerium})  & \texttt{BERT-base} & $128$ & $43.6\rlap{$^{\S}$}$ & {\scshape RTX PRO 6000} & \yes \\
      \rowcolor{RoyalBlue!12}\textbf{\texttt{Perseus} (ours)} & \texttt{BERT-base} & $128$ & $\mathbf{8.6}$ & {\scshape RTX PRO 6000} & \yes \\
      \bottomrule
    \end{NiceTabularX}
    \label{tab:bert}
  \end{threeparttable}
\end{table}

\vit\texttt{ViT} and\bert\texttt{BERT}, follow the prefill implementation, with the only difference in Attention patterns, and we test respectively encrypted sentence and image classification (\cref{tab:bert,tab:vit}). We achieve state-of-the-art latency in both cases, proving \texttt{Perseus}' efficiency and coverage of modern deep learning tasks. Their normalized accuracies are $100\%$ on $64$ {\scshape SST-2} validation sentences for\bert\texttt{BERT} and $96.9\%$ on $64$ \texttt{Tiny-ImageNet} validation images for\vit\texttt{ViT}.

\section{Related Works}
{\scshape CKKS} naturally supports {\scshape SIMD} operations, a parallelism for which modern {\scshape GPU}s are built and keep getting better at. That's why, even in the encrypted landscape, classical {\scshape CPU} frameworks such as \texttt{OpenFHE}~\cite{openfhe}, \texttt{SEAL}~\cite{seal}, and \texttt{Lattigo}~\cite{lattigo} are losing ground in favor of backends with {\scshape GPU} support~\cite{fideslib,phantom,heongpu,tensorfhe}, including the closed-source \texttt{HEaaN}~\cite{heaan} on which current state-of-the-art encrypted latencies are reported~\cite{sylph}. For this reason newest approaches~\cite{cheddar,theodosian} moved encrypted computation to $32$-bit arithmetic, aligning to {\scshape GPU}s' cores design. \texttt{Medha}~\cite{medha} and \texttt{REED}~\cite{reed} halve key storage by regenerating the random part of each key-switching key from a seed, but only in custom hardware, while \texttt{Perseus} regenerates it in-kernel on the {\scshape GPU}.

Encrypted models have scaled too. Thanks to more mature systems and faster hardware, the field is shifting from smaller models such as \texttt{BERT}~\cite{nexus,thor,moai} towards bigger {\scshape LLM}s~\cite{encryptedllm,cachemir,sylph,cerium}, yet most of these systems usually improve only one level of the stack and release no code or only part of it. At the (ii) data level, \texttt{STIP}~\cite{stip} and \texttt{THEMIS}~\cite{themis} already use the imaginary half of the slots, but only within the prefill stage, while our complex-\texttt{Cachemir} brings it to decoding. At the (iii) model level, existing placers~\cite{dacapo,fhelipe,orion,resbm,orbit,arhe} decide from depth, latency, and scale, none from the values themselves, while ours also reads the coefficient magnitude and packing period of every ciphertext.

\section{Conclusions}
We presented \texttt{Perseus}, a publicly available Python library for encrypted model inference. Its main contribution is state-of-the-art latency across multiple models and modalities, up to practically useful model sizes. It introduces optimizations at all levels of the stack, namely key regeneration, complex-\texttt{Cachemir}, and magnitude-aware placement. We also proposed a new workbench, a public speedrun on encrypted \texttt{GPT-2}, where the community's optimizations can compound.

\newpage
\section*{Acknowledgments}
This work was supported through the MUR FIS2 grant n.\ FIS-2023-00942 “NEXUS” (CUP B53C25001030001), and by NSF award \#2439951. We thank Cuiyi Zhang for her valuable advice on the choice of software license.

\printbibliography

\end{document}